\documentclass{article}
\usepackage {graphicx}

\begin{document}
\begin{center}
{\large \bf A Lithium Experiment in\\
the Program of Solar Neutrino Research.}

\vskip 0.2in

A.Kopylov, I.Orekhov, V.Petukhov, A.Solomatin\\
Institute of Nuclear Research of Russian Academy of Sciences
\end{center}

\begin{abstract}
The experiments sensitive to pp-neutrinos from the Sun are very
perspective for the precise measurement of a mixing angle $\theta
_{12}$. A $\nu $e$^{-}$ scattering experiment (Xmass) and/or a
charged-current experiment (the indium detector) can measure the
flux of electron pp-neutrinos. One can find the total flux of
pp-neutrinos from a luminosity constraint after the contribution of
$^7$Be and CNO neutrinos to the total luminosity of the Sun are
measured. The radiochemical experiment utilizing a lithium target
has the high sensitivity to the CNO neutrinos, thus, it has a good
promise for the precise measurement of a mixing angle and for the
test of a current theory of the evolution of the stars.
\end{abstract}

The main issue addressed by a currently developed lithium detector
is the measurement of the contribution of a CNO cycle to the total
luminosity of the Sun. According to a Standard Solar Model \cite{1}
this should be on the level of 1\%. The results of gallium
experiments \cite{2} give the limits less than 3.5\% (on a 1$\sigma
$ level) and 7\% (on a 3$\sigma $ level). Lithium experiment is
potentially capable to achieve the level of 0.5\% (1$\sigma $)
\cite{3}. Why it is important to reach this level of accuracy?
According to a solar model \cite{4} during the evolution of the Sun
the concentration of $^{12}$C in the central region of the Sun has
been diminished as a result of a burning it in a CNO cycle while the
concentration of $^{14}$N has been increased as a product of a CNO
cycle, see Fig.1. It is very interesting to test this
experimentally. The accurate measurement of the fluxes of the CNO
neutrinos will provide the answer to this question. This will be
useful also as a general test of a current theory of the evolution
of the stars.

Another issue is a more accurate determination of a mixing angle
$\theta _{12}$. Of the two parameters of neutrino oscillations
$\Delta $m$_{12}^2$ and $\theta _{12}$ the KamLAND experiment
\cite{5} measured the first one using the reactor antineutrinos
while the second has been found mainly upon the results of the solar
neutrino experiments \cite{6}--\cite{11}. The important point is
that for KamLAND both signs of $\Delta $m$_{12}^2$ are acceptable
while the results of the solar neutrino experiments are compatible
only with a positive sign (the mass eigenstate with the larger
electron neutrino component has a smaller mass). The determination
of the sign of $\Delta $m$_{12}^2$ would be a great fundamental
result because this will be the first step in the identifying a
neutrino mass hierarchy. The current interpretation of the results
of the solar neutrino experiments is heavily dependent on a very
essential point -- that our understanding of the neutrino
oscillations in matter (MSW effect \cite{12}) is correct in a sense
that there are no missing points in the currently accepted
description of this process. Although the agreement of all data in
the framework of this consideration is very impressive, the general
opinion is that still it requires some manifestation in a direct
experiment \cite{13}. This is why we need more data on neutrino
oscillations. One of the ways would be to measure with a very high
precision a mixing angle for reactor antineutrinos and for solar
neutrinos. If there is any even a tiny difference it will denote
unambiguously that something is missing in our present understanding
of neutrino oscillations in matter. The main reason why Kamland
turned out to be rather limited in the determination of a mixing
angle is not an optimal distance from a reactor \cite{14}.

\begin{figure}[!h]
\centering
\includegraphics[width=3in]{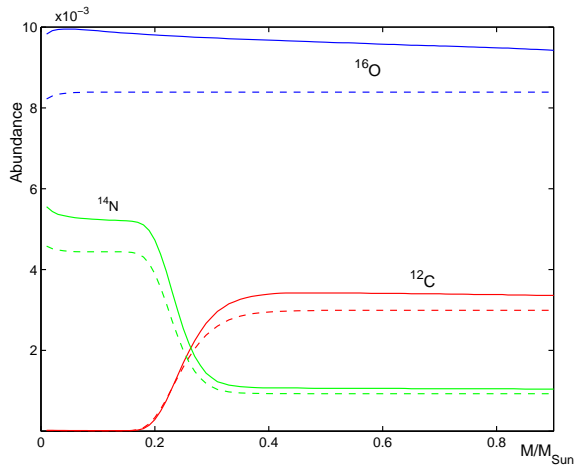}
\caption{The abundances of $^{12}C$ and $^{14}N$ in the interior of
the Sun. The solid lines - with diffusion in the matter of the Sun,
the dashed lines - without diffusion.}
\end{figure}

\noindent One can obtain much higher accuracy for a mixing angle if
to place the detector in the point of a maximal oscillation effect
(minimum in the oscillation curve). The (SADO) detector \cite{15}
located in Mt. Komagatake at the distance 54 km from a reactor
complex Kashivazaki-Kariva NPP in Japan can reach a world-record
accuracy on $sin^2\theta _{12} \approx 2\% (\approx 3\%)$ at 68.27\%
CL by 60 GW$ \cdot $kt$ \cdot $yr (20 GW$ \cdot $kt$ \cdot $yr)
operation. The important point also is that this experiment (SADO)
would provide a direct prove that the minimum of the oscillation
curve really exist. The present data of KamLAND are very compelling
to accept this fact, but it would be good to have a direct
experimental prove. Is there any possibility to achieve a similar
sensitivity in the solar neutrino experiments? Let us look at the
results of SNO experiment \cite{11}. From the ratio of the flux of
electron neutrinos (found from a charged-current interaction) to the
total flux of neutrinos from the Sun (found from a neutral-current
interaction) one obtained the value of $\theta _{12}$. The fluxes
are determined with the statistical uncertainties of about 4\% and
systematic ones of about 5 \%. The current uncertainties in the
determination of $\theta _{12}$ = $33.9_{-2.2}^{+2.4}$ are mainly
limited by these numbers.

\begin{figure}[!ht]
\centering
\includegraphics[width=2.2in]{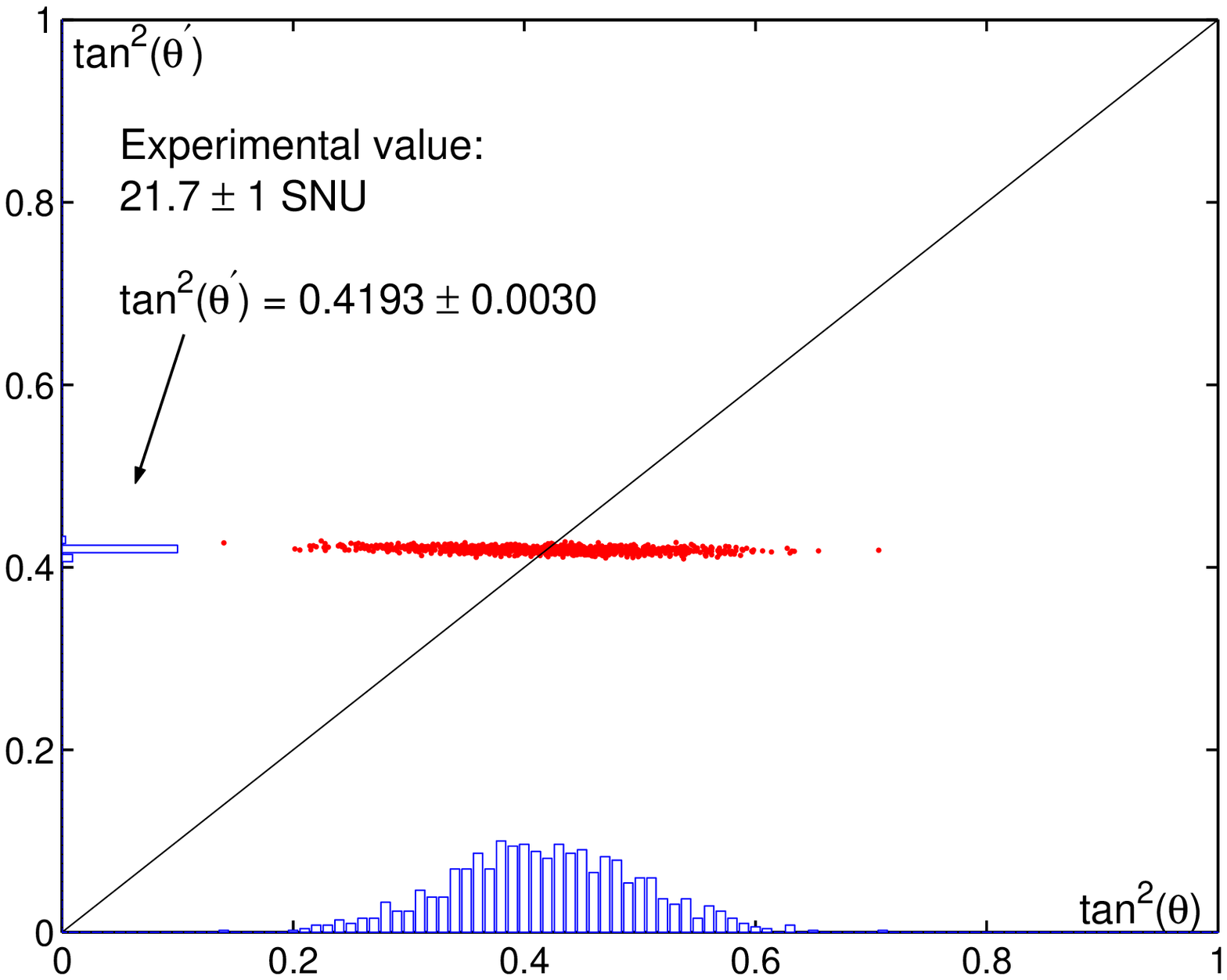}
\includegraphics[width=2.2in]{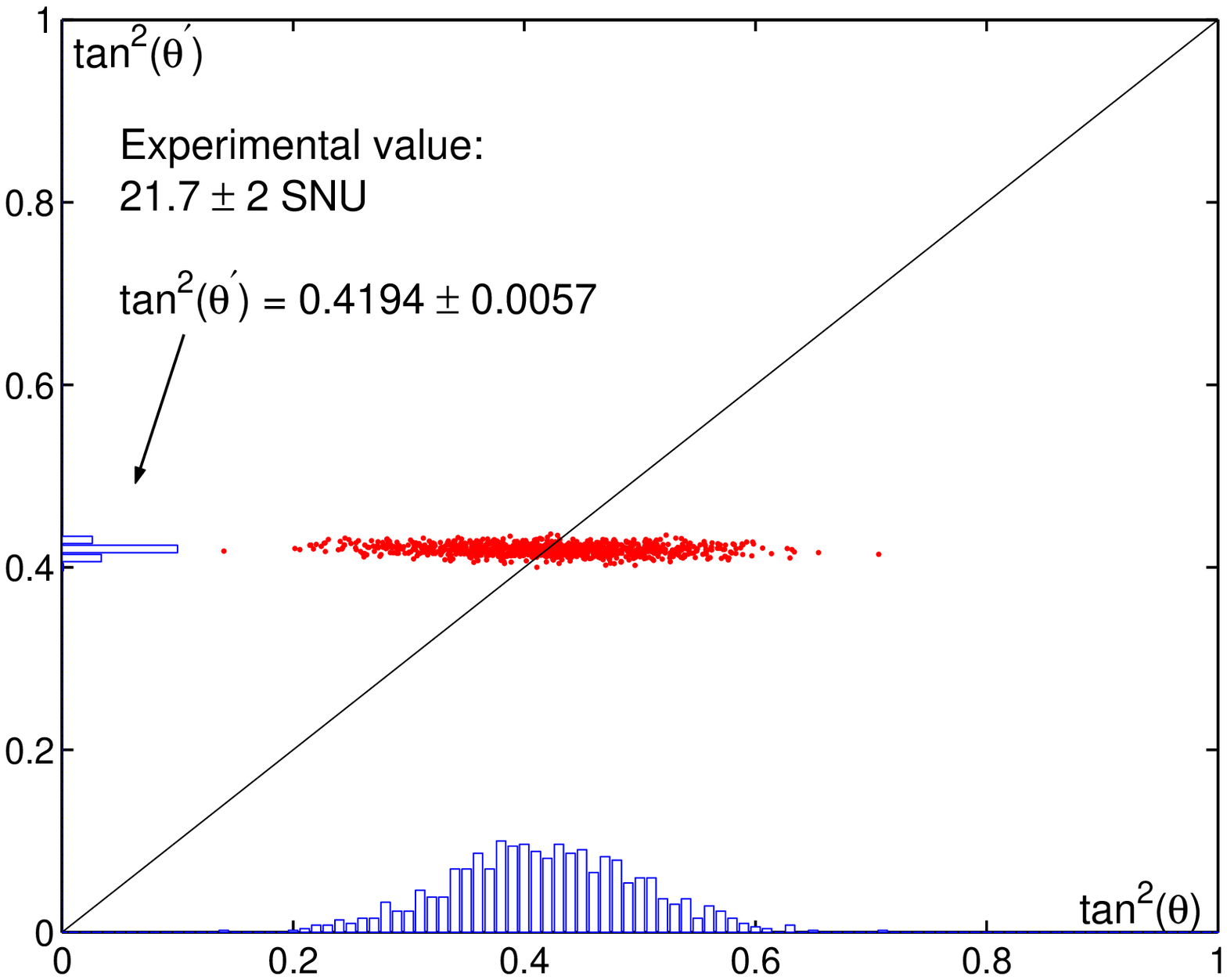}
\includegraphics[width=2.2in]{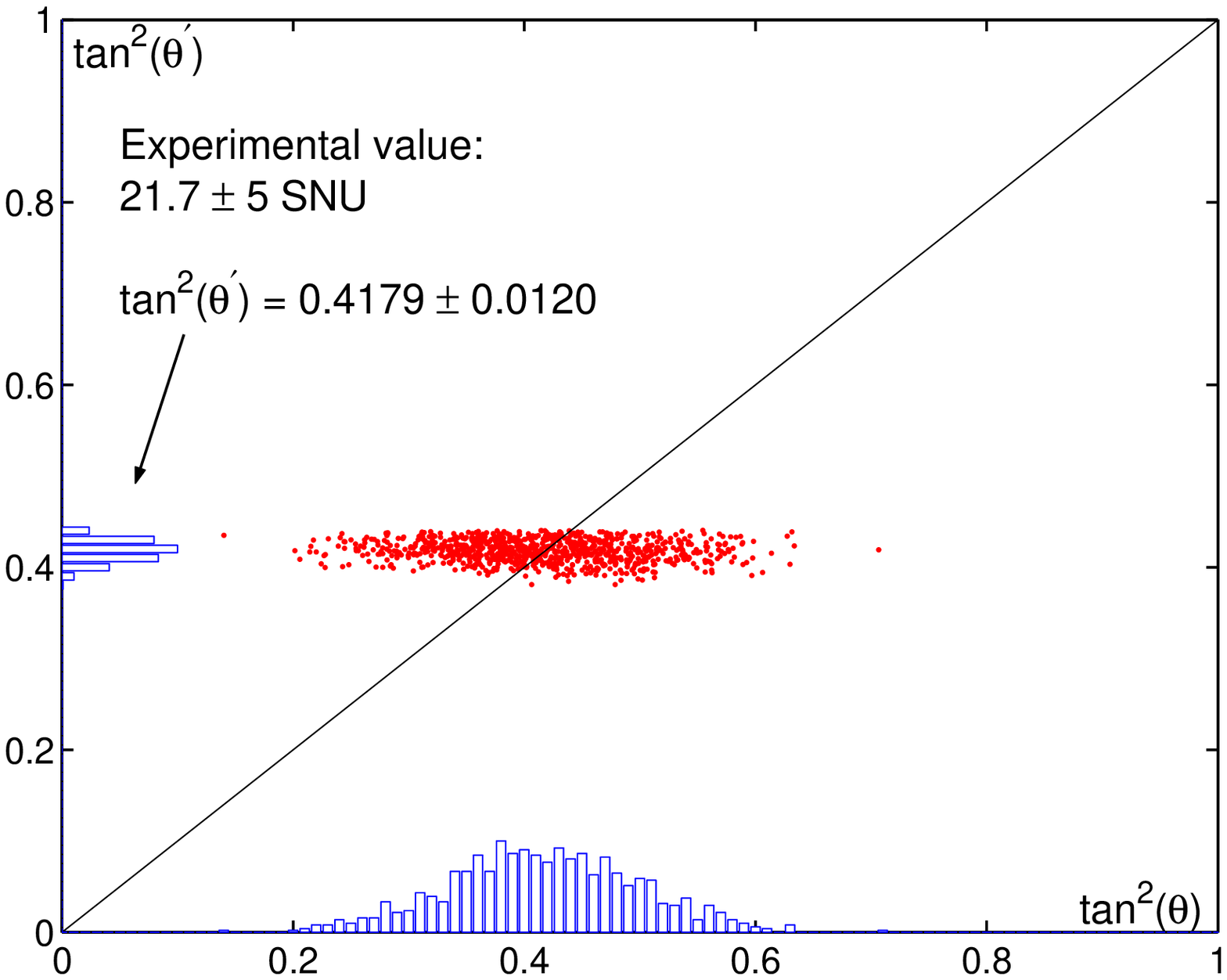}
\includegraphics[width=2.2in]{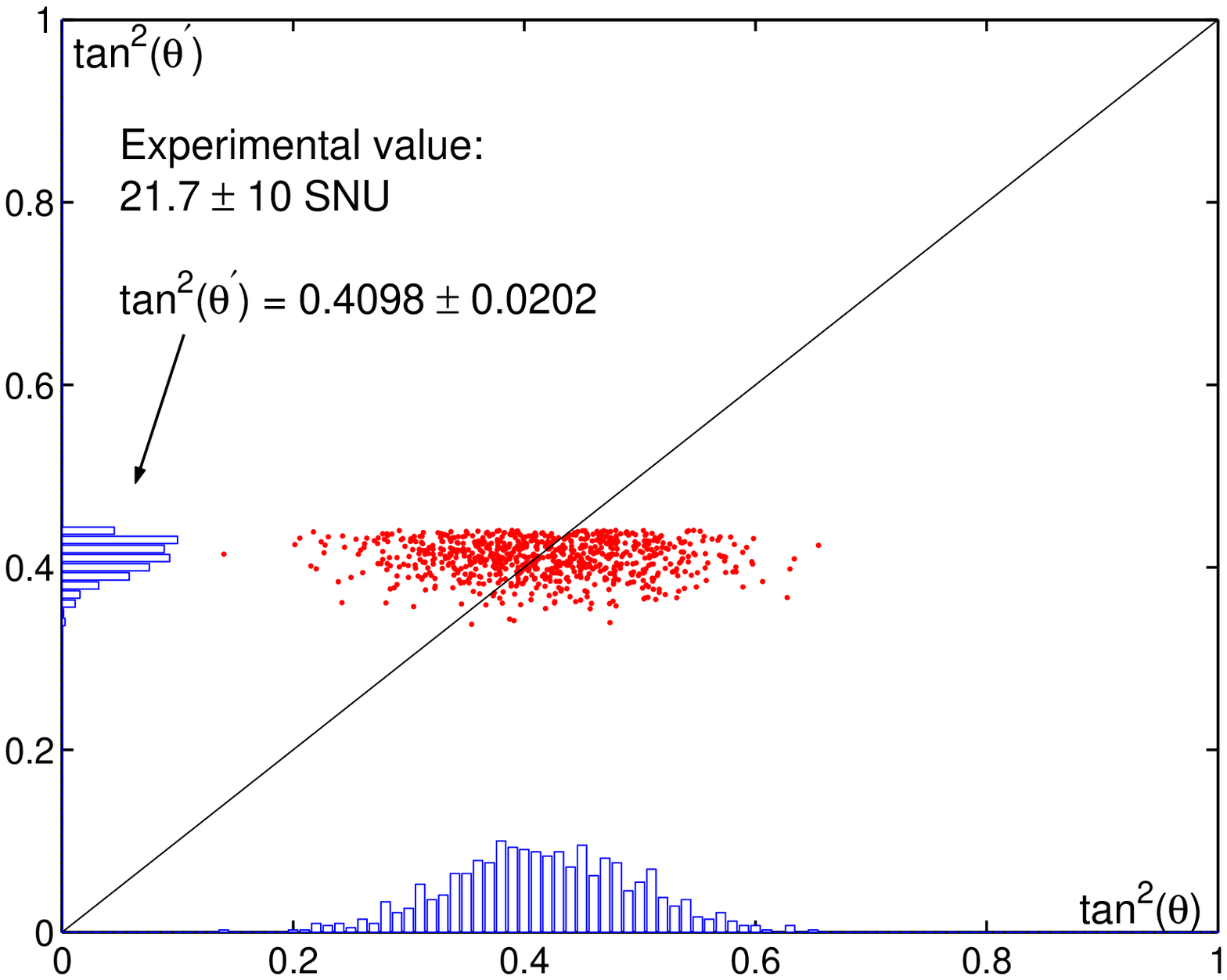}
\caption{The accuracy of the evaluation of $\theta^{\prime}$ for
different uncertainties of a lithium experiment. The distribution on
Y-axis is obtained with the Luminosity Constraint applied to the
data.}
\end{figure}

\noindent One can get a further increase of accuracy by means of a
substantial improvement in the measurement of the neutrino fluxes.
There are several possibilities, which are under development now.
Xmass project \cite{16} is trying to utilize the reaction of $\nu
$e$^{-}$ scattering on Xe. Here one attempts to observe a clear
signal from the low energy pp neutrinos. The aim is to reach the
accuracy of about 1-2\% for the flux of electron pp neutrinos.
Another approach suggested by R.Raghavan \cite{17} is to use a
charged-current interaction, which can give a clear signal for the
electron pp-neutrinos. Because the cross-section of the neutrino
capture reaction by a proposed $^{115}$In target has large
uncertainties, this will need a very powerful neutrino source to
calibrate the detector. Both projects have the aim to measure the
flux of electron pp neutrinos to improve the accuracy in
determination of a mixing angle. Following the strategy realized
in SNO, it would be very useful to know the total flux of pp
neutrinos also. We suggest a non-standard approach. Instead of
measuring directly the total flux of pp-neutrinos, we propose to
determine the flux from a luminosity constraint by evaluating the
contribution of non-pp neutrinos ($^7$Be and CNO). If to measure
the flux of $^7$Be neutrinos with the accuracy of about 10\% and
the flux of CNO neutrinos -- with the accuracy of about 30\% this
will enable to determine the total flux of pp-neutrinos with the
accuracy of about 1\% \cite{3}. The low weights of the neutrino
generating thermonuclear reactions in the total luminosity of the
Sun explain the high tolerant uncertainties for $^7$Be and CNO
neutrinos. Thus by measuring the flux of CNO neutrinos, a lithium
experiment will give an essential ingredient -- the total flux of
pp-neutrinos generated in the Sun, which will enable to improve
the accuracy in the determination of a mixing angle. A nontrivial
moment in this consideration is the following. Lithium detector is
measuring the flux of electron neutrinos coming to the Earth, i.e.
with the attenuation factor due to the oscillation effect. This
factor is the function of a mixing angle. As the input data we use
a mixing angle with the presently obtained uncertainties. After
the evaluation of the contribution of CNO cycle to a total
luminosity of the Sun, we can find precisely the total flux of
pp-neutrinos. (Here we assume that by the time a lithium
experiment collects data Borexino and KamLAND will measure the
flux of $^7$Be neutrinos with the accuracy of at least 10\% what
will enable to find the contribution of the $^7$Be neutrino
generated reactions to a total luminosity of the Sun with the
uncertainty $\textless$ 1\%.) Then by comparing the flux of
pp-neutrinos obtained from the data of $\nu $e$^{-}$ scattering
experiment (Xmass) with the total flux of pp-neutrinos we can find
precisely a mixing angle as the result on the output. This
iterative procedure becomes possible because the weight of the CNO
cycle to the total luminosity of the Sun is small (about 1\%), so
that the input uncertainties of a mixing angle (about 10\%) weakly
interfere with the output uncertainties (a few percent), see
Fig.2. We presented the results of the calculation according to
this scheme in \cite{18} where we described the detailed procedure
of the calculation.

G.Zatsepin, V.Kuzmin \cite{19} and J.Bahcall \cite{20} were first
who proposed a lithium detector. Lithium has a high and precisely
calculated cross-section for the neutrino capture reaction because
the transition is super allowed. First shown by G.Domogatsky
\cite{21} and later calculated more precisely by J.Bahcall
\cite{22} is a non-trivial effect, that due to a thermal
broadening in the central region of the Sun the $^7$Be neutrinos
contribute significantly to the total yield of $^7$Be in a lithium
target. One can see the production rates for lithium in Table 1.

\begin{table}[!h]
\caption{ Standard Model Predictions (BP2000): solar neutrino fluxes
and neutrino capture rates without neutrino oscillations, with
1$\sigma$ uncertainties from all sources (combined quadratically).}
\begin{tabular}
{|c|c|c|c|c|} \hline Source& Flux \par (10$^{10}$cm$^{-2}$s$^{-1}$)&
Cl
\par (SNU)& Ga \par (SNU)& Li \par (SNU) \\ \hline pp&
5.95(1.00$^{+0.01}_{-0.01}$)& 0.0& 69.7& 0.0
\\ \hline pep& 1.40$\times $10$^{-2}$(1.00$^{+0.015}_{-0.015}$)& 0.22& 2.8& 9.2
\\ \hline hep& 9.3$\times $10$^{-7}$& 0.04& 0.1& 0.1
\\ \hline $^{7}$Be& 4.77$\times $10$^{-1}$(1.00$^{+0.10}_{-0.10}$)& 1.15& 34.2& 9.1
\\ \hline $^{8}$B& 5.05$\times $10$^{-4}$(1.00$^{+0.20}_{-0.16}$)& 5.76& 12.1& 19.7
\\ \hline $^{13}$N& 5.48$\times $10$^{-2}$(1.00$^{+0.21}_{-0.17}$)& 0.09& 3.4& 2.3
\\ \hline $^{15}$O& 4.80$\times $10$^{-2}$(1.00$^{+0.25}_{-0.19}$)& 0.33& 5.5& 11.8
\\ \hline $^{17}$F& 5.63$\times $10$^{-4}$(1.00$^{+0.25}_{-0.25}$)& 0.0& 0.1& 0.1
\\ \hline Total& & 7.6$^{+1.3}_{-1.1}$& 128$^{+9}_{-7}$& 52.3$^{+6.5}_{-6.0}$
\\ \hline
\end{tabular}
\end{table}

By the total production rate 25 SNU expected in case of neutrino
oscillation with the presently obtained parameters, 8 SNU is the
contribution of neutrinos from a CNO cycle, i.e. approximately 30\%
while the contribution of a CNO cycle to the total luminosity of the
Sun is only 1.5\% . One studied a radiochemical lithium detector as
a perspective one since long ago \cite{23}. The main problems were
the difficulties in the extraction procedure of $^7$Be and in the
counting of a few atoms of $^7$Be extracted from the target. The
present status of the work on a lithium project is the following.
The work on the laboratory installations enabled to choose the
optimal layout of the installation: it has a modular structure -- 20
modules, each module contains 500 kg of lithium. By now, we
completed the development of the technology of extraction of
beryllium from lithium on the laboratory installations. We are
planning to perform the counting of $^7$Be by means of the array of
HPGe detectors. Now the work is going on the construction of the
first module of the full-scale lithium detector as a pilot lithium
installation. The aim is to demonstrate on one module the effective
functioning of the technique.

This work was supported in part by a Russian Foundation for Basic
Research (project no. 04-02-16678), by the grant of Russia Leading
Scientific Schools LSS-1786.2003.2 and by the Program of
fundamental research of Presidium of Russian Academy of Sciences
"Neutrino Physics". The authors deeply appreciate the very
stimulating discussions with G.Zatsepin, L.Bezrukov and V.Kuzmin
and express a gratitude to B.Zhuikov for the help in the
irradiation of an aluminum sample in a proton beam of a Moscow
meson facility which enabled us to complete the work on the
development of the technology of the beryllium extraction from a
lithium target.

\end{document}